\documentclass{appolb}

\usepackage{graphicx}
\usepackage{subfigure}
\usepackage{cite}

\begin{document}

\title{Flow in p-Pb collisions at the LHC%
\thanks{Presented by WB at {\em Excited QCD 2013}, 3-9 February 2013, Bjelasnica Mountain, Sarajevo}%
}

\author{Wojciech Broniowski
\address{The H. Niewodnicza\'nski Institute of Nuclear Physics PAN, 31-342 Cracow, Poland\\
and Institute of Physics, Jan Kochanowski University, 25-406~Kielce, Poland}
\and
Piotr Bo\.zek
\address{AGH University of Science and Technology, Faculty of Physics and Applied Computer Science, al. Mickiewicza 30, 30-059 Krakow, Poland\\
and The H. Niewodnicza\'nski Institute of Nuclear Physics PAN, 31-342 Cracow, Poland}
}

\maketitle

\begin{abstract}
We present the predictions of a hydrodynamic model for the flow observables recently measured in the 
highest-multiplicity p+Pb collisions at the LHC. We focus o the ``ridge'' phenomenon, which provides an important probe 
of the long-range dynamics and may be used to support the collective interpretation of the p+A data. 
\end{abstract}

 
\bigskip
\bigskip
 
In this talk we review our recent predictions~\cite{Bozek:2012gr,Bozek:2013df,Bozek:2013uha} concerning the 
possibility of soft collective dynamics in the highest-multiplicity p+Pb collisions an the LHC energies of $\sqrt{s_{NN}}=5.02$~TeV. 
Indeed, one of the most important findings of the heavy-ion program at RHIC, now confirmed at the LHC, is the collective flow 
appearing in the A+A collisions~\cite{Ollitrault:1992bk}. It results, inter alia, in the {\em ridge} phenomenon in the two-particle two-dimensional 
correlations in the relative azimuth and pseudorapidity in the high multiplicity A+A 
collisions~\cite{Collaboration:2011pfa,Chatrchyan:2013nka,Aamodt:2011by,Adare:2008ae,Abelev:2009af,Wenger:2008ts}, as well as
in the {\em highest} multiplicity p+A collisions and even the p+p collisions~\cite{Khachatryan:2010gv,Adare:2013piz,Aad:2012gla,Chatrchyan:2013nka,Abelev:2012ola}.

The appearance of the ridge in A+A collisions found a convincing explanation in terms of the collective harmonic flow~\cite{Takahashi:2009na,Alver:2010gr}.
Indeed, if the created fireball in the transverse plane is approximately boost-invariant, i.e., if the resulting flow patterns in the transverse 
directions are similar over a range of pseudorapidity (typically, experiments cover from a few units), then a collimation effect appears, extending over a few 
units of rapidity. One may imagine surfers on a long wave: they all move in the same directions due to the flow, even if they are two miles away!

The pertinent question now is if the flow idea may also be applied to a much smaller system than the one formed in central A+A collisions, 
namely, the fireball in p+Pb collisions. We use a three-stage approach which became popular in the A+A studies, where it describes successfully 
numerous aspects of the soft-physics data. The three phases are:

\begin{enumerate}
 
\item The fluctuating initial state,
obtained here with the Glauber simulations~\cite{Broniowski:2007nz}, where the initial 
density is obtained by placing smeared sources in the centers of the 
participating nucleons.

\item The subsequent event-by-event hydro simulations with the 3+1D viscous dynamics~\cite{Bozek:2011if,Bozek:2013uha}, 
with the shear viscosity $\eta/s=0.08$, the bulk viscosity, and the early hydro ignition time of $\tau=0.6$~fm. 

\item The statistical hadronization~\cite{Chojnacki:2011hb}, carried out  
at the constant freeze-out temperature $T_f-150$~MeV. 

\end{enumerate}

The physical object of our study is the per-trigger correlation function in relative pseudorapidity and azimuth, defined as~\cite{CMS:2012qk}
\begin{equation}
\label{eq:2pc} \hspace{-4mm}
C_{\rm trig}(\Delta\eta,\Delta\phi) \equiv \frac{1}{N}\frac{d^{2} N^{\rm{pair}}}{d\Delta\eta\, d\Delta\phi}
= B(0,0) \frac{S(\Delta\eta,\Delta\phi)}{B(\Delta\eta,\Delta\phi)},
\end{equation}
where $\Delta\eta$ and $\Delta\phi$ are the relative pseudorapidity and azimuth
of the particles in the pair. The signal is defined via the pairs from the same event,
\begin{equation}
\label{eq:s}
S(\Delta\eta,\Delta\phi) = \langle \frac{1}{N}\frac{d^{2}N^{\rm{same}}}{d\Delta\eta\, d\Delta\phi} \rangle_{\rm events} ,
\end{equation}
whereas the mixed-event background distribution is
\begin{equation}
\label{eq:b}
B(\Delta\eta,\Delta\phi) = \langle \frac{1}{N}\frac{d^{2}N^{\rm{mix}}}{d\Delta\eta\, d\Delta\phi} \rangle_{\rm mixed \ events}.
\end{equation}
The variable $N$ denotes the number of charged particles in 
a given centrality class and acceptance bin. 
To make quantitative comparisons, one introduces the projected correlation function
\begin{eqnarray}
 Y(\Delta \phi)=\frac{\int B(\Delta \phi) d(\Delta \phi)}{\pi N} C(\Delta\phi) - b_{\rm ZYAM}, \label{eq:Y}
\end{eqnarray}
where $S(\Delta \phi)$ and $B(\Delta \phi)$ are averages of $S(\Delta \eta, \Delta \phi)$ and $B(\Delta \eta, \Delta \phi)$ over the 
chosen range in $\Delta \eta$ avoiding the central region, in particular $2< |\Delta \eta| < 5$ in the ATLAS analysis, and the constant 
$b_{\rm ZYAM}$ is such that the minimum of $Y(\Delta \phi)$ is at zero.

\begin{figure}[tb]
\begin{center}
\includegraphics[width=.799\textwidth]{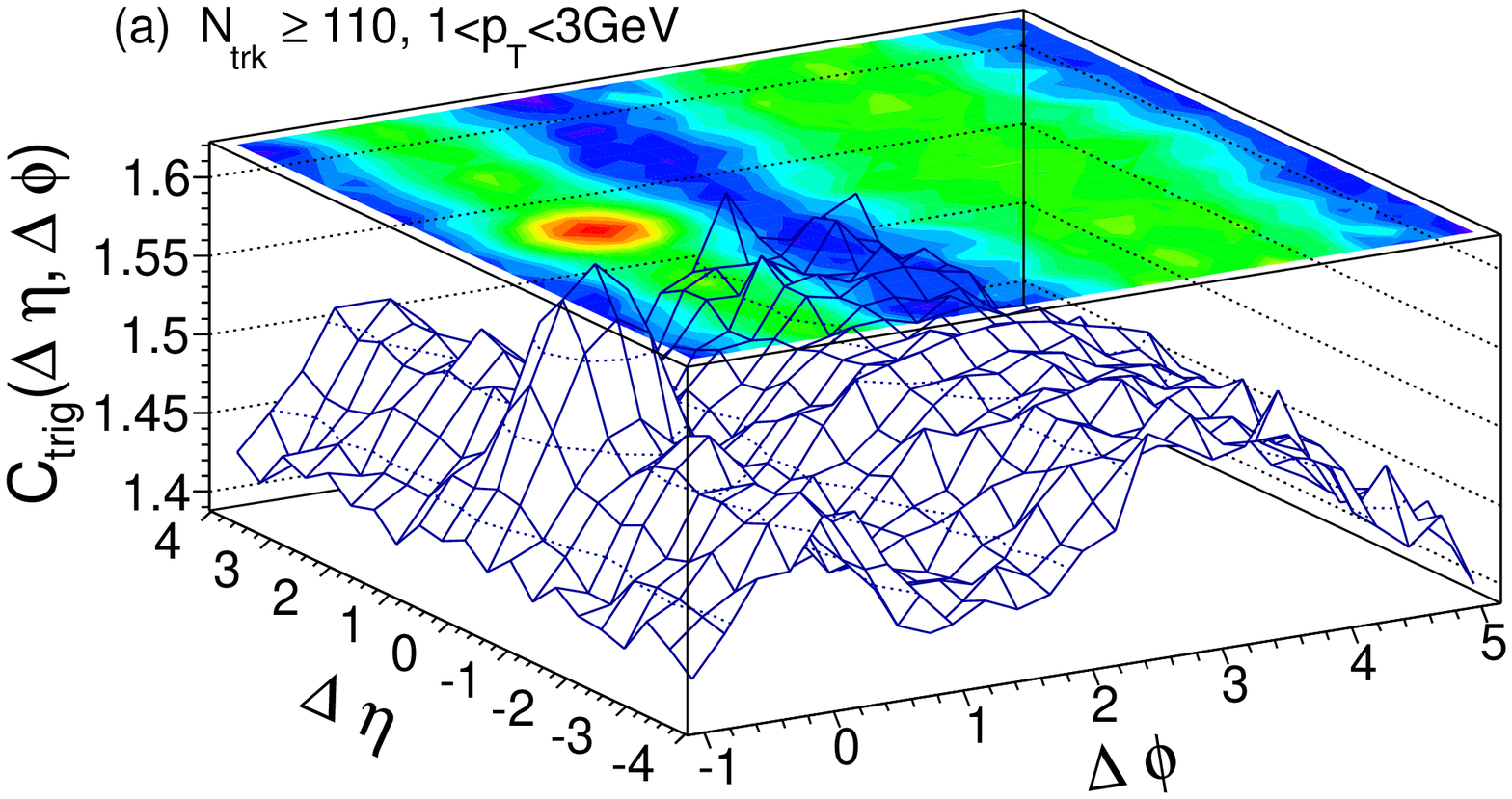}
\includegraphics[width=.799\textwidth]{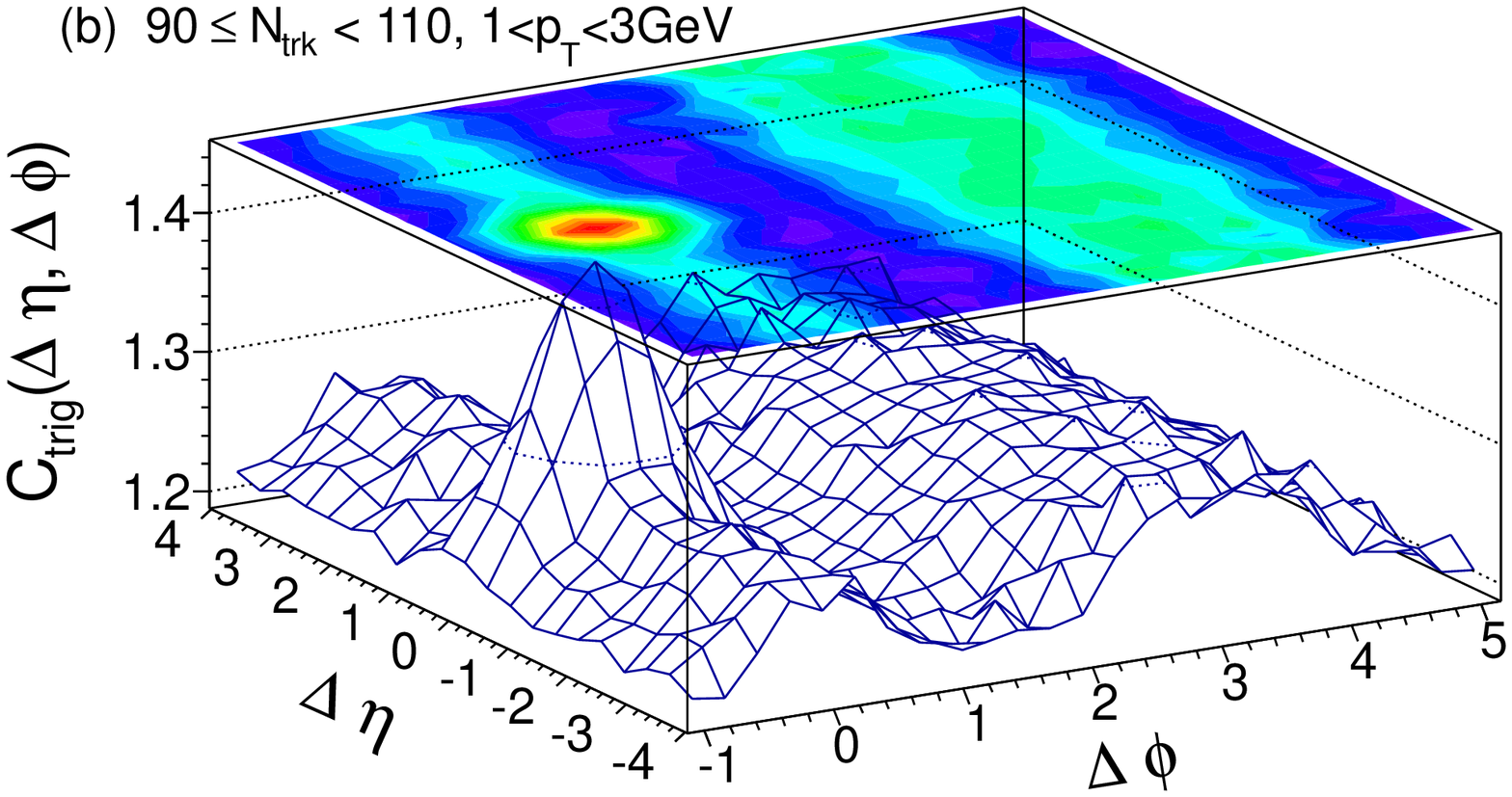}
\end{center}
\caption{The per-trigger-particle 
correlation function $C_{\rm trig}(\Delta\eta, \Delta\phi)$ of Eq.~({\ref{eq:2pc}}) for the two most central centrality classes 
corresponding to $N_{track} \ge 110$  (panel a) and $90 \le N_{track} \le 110$ (panel b) 
for the p+Pb collisions used by the CMS Collaboration. The transverse momentum of each particle belongs to 
the relatively soft range $1.0 < p_{T} < 3.0$~GeV. \label{fig:ridge}}
\end{figure}

\begin{figure}[tb]
\begin{center}
\includegraphics[width=.8\textwidth]{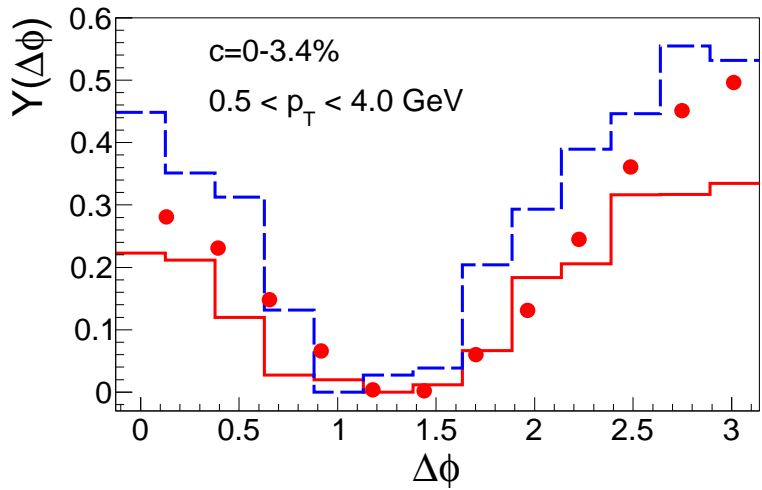}
\end{center}
\caption{The projected and ZYAM-subtracted correlation function $Y\Delta(\phi)$ for the most central p-Pb collisions
for the standard source (solid line) and compact source (dashed line), compared to the most central ATLAS data (points). 
The total transverse momentum is approximately conserved with the condition for the total transverse momentum, $P_T<5$~GeV. 
Charge balancing is imposed. \label{fig:zyam}}
\end{figure}

In Fig.~\ref{fig:ridge} we show the typical results of our simulations. We note the two ridges, as well as the central peak, here formed due 
to included charge balancing~\cite{Bozek:2012is}, and to a lesser extent by the decays of 
resonances. In Fig.~\ref{fig:zyam} we show the  projected and ZYAM-subtracted correlation function for two variants of the model, with the standard 
and compact sources (cf. Ref.~\cite{Bozek:2013uha}). We note a fair agreement of the hydrodynamical model with the data. 
Other results, such as the obtained harmonic flow coefficients, are given in Ref.~\cite{Bozek:2013uha}.

In conclusion, we note that the correlation and flow data for the highest multiplicity p+Pb collisions at the LHC may be satisfactorily described with the 
approach incorporating hydrodynamics, thus based on collectivity of the dynamics. Thus we offer an explanation following the path of the A+A analyses. 
For scenarios based on the the Color Glass Condensate theory see~\cite{Dusling:2012iga,Dusling:2012wy,Dusling:2012cg,Kovchegov:2012nd,Kovner:2012jm,Dusling:2013oia}.
The discrimination of the two approaches may be made in the future with the particle-identified data.

\bigskip
We gratefully acknowledge the support of the Polish
National Science Centre, grant DEC-2012/06/A/ST2/00390, and PL-Grid infrastructure.

\bibliography{hydr}

\begin{thebibliography}{10}

\bibitem{Bozek:2012gr}
P. Bo\.zek and W. Broniowski,
\newblock Phys. Lett. B718 (2013) 1557, 1211.0845.

\bibitem{Bozek:2013df}
P. Bo\.zek and W. Broniowski,
\newblock Phys. Lett. B720 (2013) 250, 1301.3314.

\bibitem{Bozek:2013uha}
P. Bozek and W. Broniowski,
\newblock Phys. Rev. C 88, 014903 (2013), 1304.3044.

\bibitem{Ollitrault:1992bk}
J.Y. Ollitrault,
\newblock Phys. Rev. D46 (1992) 229.

\bibitem{Collaboration:2011pfa}
ATLAS Collaboration, S. Mohapatra,
\newblock J.Phys.Conf.Ser. 389 (2012) 012011, 1109.6721.

\bibitem{Chatrchyan:2013nka}
CMS Collaboration, S. Chatrchyan et~al.,
\newblock (2013), 1305.0609.

\bibitem{Aamodt:2011by}
ALICE Collaboration, K. Aamodt et~al.,
\newblock Phys.Lett. B708 (2012) 249, 1109.2501.

\bibitem{Adare:2008ae}
PHENIX Collaboration, A. Adare et~al.,
\newblock Phys.Rev. C78 (2008) 014901, 0801.4545.

\bibitem{Abelev:2009af}
STAR Collaboration, B. Abelev et~al.,
\newblock Phys.Rev. C80 (2009) 064912, 0909.0191.

\bibitem{Wenger:2008ts}
PHOBOS Collaboration, B. Alver et~al.,
\newblock J.Phys. G35 (2008) 104080, 0804.3038.

\bibitem{Khachatryan:2010gv}
CMS, V. Khachatryan et~al.,
\newblock JHEP 09 (2010) 091, 1009.4122.

\bibitem{Adare:2013piz}
PHENIX Collaboration, A. Adare et~al.,
\newblock (2013), 1303.1794.

\bibitem{Aad:2012gla}
ATLAS Collaboration, G. Aad et~al.,
\newblock Phys.Rev.Lett. 110 (2013) 182302, 1212.5198.

\bibitem{Abelev:2012ola}
ALICE Collaboration, B. Abelev et~al.,
\newblock Phys.Lett. B719 (2013) 29, 1212.2001.

\bibitem{Takahashi:2009na}
J. Takahashi et~al.,
\newblock Phys. Rev. Lett. 103 (2009) 242301, 0902.4870.

\bibitem{Alver:2010gr}
B. Alver and G. Roland,
\newblock Phys. Rev. C81 (2010) 054905, 1003.0194.

\bibitem{Broniowski:2007nz}
W. Broniowski, M. Rybczy\'nski and P. Bo\.zek,
\newblock Comput. Phys. Commun. 180 (2009) 69, 0710.5731.

\bibitem{Bozek:2011if}
P. Bo\.zek,
\newblock Phys. Rev. C85 (2012) 014911, 1112.0915.

\bibitem{Chojnacki:2011hb}
M. Chojnacki et~al.,
\newblock Comput. Phys. Commun. 183 (2012) 746, 1102.0273.

\bibitem{CMS:2012qk}
CMS Collaboration, S. Chatrchyan et~al.,
\newblock Phys. Lett. B718 (2013) 795, 1210.5482.

\bibitem{Bozek:2012is}
P. Bozek and W. Broniowski,
\newblock Nucl.Phys.A904-905 2013 (2013) 479c, 1210.4315.

\bibitem{Dusling:2012iga}
K. Dusling and R. Venugopalan,
\newblock Phys.Rev.Lett. 108 (2012) 262001, 1201.2658.

\bibitem{Dusling:2012wy}
K. Dusling and R. Venugopalan,
\newblock Phys. Rev. D 87 (2013) 054014, 1211.3701.

\bibitem{Dusling:2012cg}
K. Dusling and R. Venugopalan,
\newblock Phys. Rev. D 87 (2013) 051502, 1210.3890.

\bibitem{Kovchegov:2012nd}
Y.V. Kovchegov and D.E. Wertepny,
\newblock Nucl.Phys. A906 (2013) 50, 1212.1195.

\bibitem{Kovner:2012jm}
A. Kovner and M. Lublinsky,
\newblock Int. J. Mod. Phys. E Vol. 22 (2013), 1211.1928.

\bibitem{Dusling:2013oia}
K. Dusling and R. Venugopalan,
\newblock Phys. Rev. D  (2013) 094034, 1302.7018.

\end{thebibliography}
\bibliographystyle{h-elsevier}

\end{document}